# Normal Fermi Surface in the Nodal Superconductor CeCoIn₅ Revealed via Thermal Conductivity

Sangyun Lee, Duk Y. Kim[*], Priscila F. S. Rosa, Eric D. Bauer,

Filip Ronning, J. D. Thompson, Shi-Zeng Lin[†], and Roman Movshovich[†]

Los Alamos National Laboratory, Los Alamos, New Mexico 87545, USA

*Permanent address: Agency for Defense Development, Daejeon 34186, Republic of Korea*

[†]*Corresponding author: S. -Z. L.: szl@lanl.gov; R. M.: roman@lanl.gov*

The thermal conductivity of heavy-fermion superconductor CeCoIn₅ was measured with a magnetic field rotating in the tetragonal *a-b* plane, with the heat current in the anti-nodal direction, $J \parallel [100]$. We observe a sharp resonance in thermal conductivity for the magnetic field at an angle $\Theta \approx 12°$, measured from the heat current direction [100]. This resonance corresponds to the reported resonance at an angle $\Theta' \approx 33°$ from the direction of the heat current applied along the nodal direction, $J \parallel [110]$. Both resonances, therefore, occur when the magnetic field is applied in the same crystallographic orientation in the two experiments, regardless of the direction of the heat current, proving conclusively that these resonances are due to the structure of the Fermi surface of CeCoIn₅. We argue that the uncondensed Landau quasiparticles, emerging with field, are responsible for the observed resonance. We support our experimental results with density-functional-theory model calculations of the density of states in a rotating magnetic field. Our calculations, using a model Fermi surface of CeCoIn$_5$, reveal several sharp peaks as a function of the field direction. Our study demonstrates that the thermal-conductivity measurement in rotating magnetic field can probe the normal parts of the Fermi surface deep inside the superconducting state.

## Introduction

Thermal conductivity is a powerful transport probe of a superconducting state. The superconducting condensate does not carry heat current; only normal quasiparticles participate in electronic (charge) thermal transport. As a result, thermal conductivity is exquisitely sensitive to the structure of the superconducting energy gap. It can easily differentiate between a conventional superconductor with a fully gapped Fermi surface (FS) with exponential dependence of thermal conductivity on temperature, and an unconventional superconductor with nodes (zeros) in the superconducting gap, which leads to a power-law temperature dependence [1, 2]. In this manner, CeCoIn$_5$ was quickly identified as an unconventional superconductor with lines of nodes (similar to high-$T_c$ cuprates) upon the discovery of superconductivity in this compound [3].

Measurements of thermal conductivity in a rotating magnetic field provide additional insight into the detailed nature of the superconducting order parameter. When measured in a rotating magnetic field, the thermal conductivity of CeCoIn$_5$ showed a 4-fold oscillation [1, 2, 4]. In this experiment, the heat current $\boldsymbol{J}$ was applied along one of the principal axes of the tetragonal plane, $\boldsymbol{J}$ ∥ [100], and the magnetic field was rotated in the tetragonal plane. The smooth oscillation was understood as a Doppler-shift effect by the magnetic field on the quasiparticles that reside in the four nodes, and the positions of the maxima identified the $d_{x^2-y^2}$ nature of the order parameter. Vorontsov and Vekhter theoretically modelled this smooth variation of thermal conductivity [5]. More realistic calculations that included the finer details of the Fermi Surface were able to explain the switch in sign of the four-fold term between the specific heat and thermal conductivity data [6], bringing consistency between both sets of data and the $d_{x2-y2}$ superconducting order parameter in CeCoIn$_5$.

In the recent lower temperature high-resolution thermal conductivity measurements with magnetic field rotating within the *a-b* plane, a resonance-like peak as a function of the field direction was observed in addition to the smooth 4-fold oscillation [7]. In that experiment the heat current $\boldsymbol{J}$ was applied along the $\boldsymbol{J}$ ‖ [110] direction, which is the nodal direction of the $d_{x2-y2}$ superconducting gap. The experimental data can be compared with the corresponding results of the theoretical calculation of thermal conductivity in a rotating field for the $d_{xy}$ order parameter and $\boldsymbol{J}$ ‖ [100] [5], with good agreement on the positions of the broad maxima of the 4-fold oscillations in the data. However, existing theories were not able to identify the origin of the sharp resonance in thermal conductivity as a function of the angle between the rotating magnetic field and the heat current direction, observed at an angle $\Theta'\approx 33°$, where $\Theta'$ is defined by the angle between $\boldsymbol{J}$ ‖ [110] and magnetic field ($\boldsymbol{H}$). To provide guidance to theory, as well as to search for the presence of similar sharp features in thermal conductivity in other geometries, here we undertake thermal conductivity measurements in rotating magnetic field in a dilution refrigerator with the heat current applied as in the original experiment of Izawa *et al.*[4], $\boldsymbol{J}$ ‖ [100].

**Thermal conductivity measurements**

A needle-like single-crystal sample (2.98×0.15×0.05 mm$^3$) was prepared with the long axis along the [100] crystallographic direction, the superconducting anti-node direction. The thermal conductivity of CeCoIn$_5$ was measured with the heat current applied along the [100] direction and with a magnetic field rotating in the tetragonal *a-b* plane, as shown in Fig. 1. The experimental apparatus and the setup are similar to those of previous reports [8, 7], but $\Theta$ is defined by the angle between $\boldsymbol{J}$ ‖ [100] and magnetic field ($\boldsymbol{H}$). Fig. 2 shows thermal conductivity data for 3 T where the sum of 2-fold and 4-fold oscillation terms well describes the overall smooth oscillation over the entire field-angle range of the measurements. The data for other applied field, from 1 T to 7 T, are presented in Fig. S1 of the supplemental material. The salient features for all values of magnetic field are the sharp resonances at ± 12° and at ± 78°, which is (90-12)°. The novel resonances can be directly connected to the 33° resonances of previous work [8], in which the heat current was applied along the [110] direction, as 33° away from [110] is the same crystallographic direction as 12° away (45-33) from the tetragonal [100] axis.

Figure 3 shows the data for all values of the applied field, broken up into two regions: -40° < $\Theta$ < 40°, which contains both ± 12° resonances [Fig. 3(a)] and $\Theta$ < -50° ($\Theta$ > +50°), which contains the -78° (78°) resonance [Fig. 3(c)]. The solid curves in Figs. 3(a,c) represent the fits to the background, with the resonance regions excluded from the fits. Figs 3(b) and 3(d) present the data after the backgrounds are subtracted, revealing the ±12° and ±78° resonances, respectively. The resonance at ±12° is most pronounced at 3 T and is suppressed in a higher magnetic field. This trend with the magnetic field is the same as the one observed for the 33° resonance in the earlier experiment (see Fig. 4) [8]. The resonance at ±78° appears to evolve slower with field, with only a slight reduction of the amplitude at 7 T.

Note that the field angle is measured from the direction of the heat current. All the resonances, ±12° and ±78° for $J$ ∥ [100] and ±33° for $J$ ∥ [110], where the angles are measured between $J$ and $H$, occur for magnetic field pointing in the same crystallographic direction of CeCoIn$_5$. The peak positions at field angles of ±12° and ±78° for the [100] crystal axis within the $a$-$b$ plane are related to each other by $C_4$ rotation and mirror reflection along the [100] or [010] direction. Our results, therefore, first demonstrate convincingly that the resonances of the thermal conductivity observed in both experiments are an intrinsic property of CeCoIn$_5$. Second, the origin of the resonance is connected to the orientation of the field within the crystal lattice and is independent of the direction of the heat current in the sample. More generally, such resonances might be a ubiquitous feature revealed in high-resolution thermal conductivity measurements of unconventional superconductors.

In general, thermal conductivity is proportional to a product of the density of states of heat carriers (DOS) and their scattering lifetime. At the observed resonances' field angles the scattering lifetime can vary significantly. The vortex line scatters quasiparticles more frequently for $\Theta \approx \pm 78°$, compared to either $\pm 33°$ or $\pm 12°$, as the vortex line becomes more perpendicular to the heat current or the vortex lattice spacing shrinks. [9] Yet, the resonance position within the crystal is independent of the heat current direction (for $\boldsymbol{J}\|[100]$ and $\boldsymbol{J}\|[110]$), which rules out the scattering lifetime as their origin. This leads us to conclude that the resonances originate from the peak in the quasiparticle density of state, and the question is what type of quasiparticles are at play. The vortex lattice can modulate the Bogoliubov quasiparticle density of state through the Doppler shift in the quasiparticle spectrum, known as the Volovik effect [10]. For the $d_{x^2-y^2}$ pairing symmetry, the DOS as a function of field angle due to the Volovik effect is maximal when the field points in the antinodal [100] and [010] directions [11]. Moreover, the feature in thermal conductivity due to the Volovik effect is generally very broad as a function of the field angle [12], in contrast to the sharp peak observed experimentally. Therefore, the Volovik effect due to Bogoliubov quasiparticles cannot explain the observed resonance at $\pm 12°$ and $\pm 78°$.

CeCoIn$_5$ is a multiband material with multiple Fermi surfaces. In the superconducting state, a superconducting gap develops on all Fermi surfaces due to interband coupling [13]. The magnitude of these gaps, however, varies and depends on the microscopic parameters of the individual Fermi surfaces. With applied magnetic fields, a weak superconducting gap on a particular Fermi surface can be suppressed through the overlapping of the vortex cores associated with the superconducting order parameter on that Fermi surface [14]. This will result in uncondensed Landau quasiparticles in the mixed state in multiband superconductors. To determine the effect of these uncondensed Landau quasiparticles on thermal conductivity in rotating magnetic field, we performed DFT calculations in the normal state of CeCoIn$_5$.

**DFT calculation**

CeCoIn$_5$ is a tetragonal system with warped cylindrical Fermi surfaces [15]. When the cross-section of the Fermi surface normal to the magnetic field direction overlaps the cyclotron orbit, a resonance condition can be satisfied, and the density of states can be enhanced. This mechanism can therefore produce a sharp enhancement of heat conduction. To shed more light on the possible origin of the resonances, we investigated the possible effects of the detailed structure of the normal Fermi surface in CeCoIn$_5$. We use the QUANTUM-ESPRESSO software [16] to compute the band structure of CeCoIn$_5$, utilizing the generalized gradient approximation (GGA) [17]. We use the Perdew-Burke-Ernzerhof functional from the standard solid-state pseudopotentials library [18]. The calculations have been performed over $32 \times 32 \times 32$ k-points. We then use the software XCrySDen to extract Fermi surfaces. Finally, we use the package SKEAF [19] to compute the density of state at the Fermi energy as a function of magnetic field angle for each Fermi surface.

CeCoIn$_5$ is a multiband superconductor, and each band is expected to have a different response to the magnetic field. The calculated density of states for three representative Fermi surfaces of CeCoIn$_5$, as the field is rotated within the $a$-$b$ plane, are shown in Fig. 5. For all of these Fermi surfaces, the density of states displays strong dependence on the direction of the magnetic field, with multiple resonant peaks present. We can thereby offer a solution to the puzzle of sharp resonances in high-resolution thermal conductivity in CeCoIn$_5$. Even though most of the Fermi surface is gapped by the superconducting $d$-wave order parameter, some weakly superconducting parts of Fermi surface become normal in the applied magnetic field [20, 21]. The normal Fermi surface then contributes to overall thermal conductivity, producing dramatic resonances in a rotating magnetic field. All Fermi surfaces develop resonances for field around $15°$ or symmetry-related angle of $75°$. Particularly, the $\alpha$ Fermi surface has the most pronounced peaks. In addition, there are smaller resonances at other angles, e.g. $30°$.

The uncondensed Landau quasiparticle origin of the resonances also naturally explains the field strength dependence of the resonance strength. Experimentally, the resonance strength increases with the field and then decreases. The initial increase is due to the suppression of the superconductivity on these Fermi surfaces and sets the stage for the Landau quasiparticles to play. At high magnetic field, superconductivity is suppressed, and the lifetime of the Landau quasiparticle reduces, which results in the reduction of the resonance strength. This may also explain why only the resonance around $15°$ is observed in the experiment because the quasiparticle lifetime is shorter for a larger angle between the field and heat current due to the scattering by the vortex lattice. For the same reason, it is likely that the resonance around 15° is mainly contributed by the $\alpha$ Fermi surface. Further efforts are required to resolve the role of individual Fermi surface.

**Discussion**

Our study demonstrates that the thermal conductivity measurement in a rotating magnetic field can provide information about the normal Fermi surface even deep in the superconducting state. Thermal conductivity is commonly considered to be a material property that varies rather slowly under changing experimental conditions, be it temperature, magnetic field, or the direction of the magnetic field. For example, let us compare the anomalies in specific heat and thermal conductivity associated with a superconducting transition in CeCoIn$_5$. Specific heat displays a jump, characteristic of the mean-field phase transition, while thermal conductivity displays only a kink at $T_c$. The expectation of a smooth, conventional variation of thermal conductivity with rotating field stands in sharp contrast with our observations of sharp resonances in thermal conductivity in a rotating magnetic field deep in the superconducting state of CeCoIn$_5$. The present results unambiguously demonstrate that this effect is intrinsic. The heat current direction – independence of the resonances links the resonances to the Fermi surface of CeCoIn$_5$. In addition to possible normal surfaces, unconventional superconductors with nodes of order parameters are expected to develop emergent Bogolyubov Fermi Surfaces (BFS) in magnetic fields. BFS can also lead to enhancement of thermal conductivity, with a potential for resonances similar to those from normal Landau Fermi Surfaces. Our results, therefore, encourage high-resolution thermal conductivity measurements in other unconventional superconductors.

**Reference**

1. Y. Matsuda, K. Izawa, and I. Vekhter, J. Phys. Condens. Matter 18, **R705** (2006).

2. H. Shakeripour, C. Petrovic, and L. Taillefer, New J. Phys. **11**, 055065 (2009).

3. R. Movshovich, M. Jaime, J. D. Thompson, C. Petrovic, Z. Fisk, P. G. Pagliuso, and J. L. Sarrao, Phys. Rev. Lett. **86**, 5152 (2001).

4. K. Izawa, H. Yamaguchi, Y. Matsuda, H. Shishido, R. Settai, and Y. Onuki, Phys. Rev. Lett. **87**, 057002 (2001).

5. A. B. Vorontsov and I. Vekhter, Unconventional Phys. Rev. B **75**, 224502 (2007).

6. P. Das et al., Phys. Rev. Lett. **108**, 087002 (2012)

7. Duk Y. Kim, Shi-Zeng Lin, Franziska Weickert, Eric D. Bauer, Filip Ronning, J. D. Thompson, and Roman Movshovich, Phys. Rev. Lett. **118**, 197001 (2017).

8. Duk Y. Kim, Shi-Zeng Lin, Franziska Weickert, Michel Kenzelmann, Eric D. Bauer, Filip Ronning, J. D. Thompson, and Roman Movshovich, Phys. Rev. X **6**, 041059 (2016).

9. I. Vekhter and H. Houghton, Phys. Rev. Lett. **83**, 22 (1999).

10. G. E. Volovik, Pis'ma Zh. Eksp. Teor. Fiz. **58**, 457 (1993).

11. I. Vekhter, P. J. Hirschfeld, J. P. Carbotte, and E. J. Nicol, Phys. Rev. B **59**, 14 (1999).

12. A. B. Vorontsov and I. Vekhter, Phys. Rev. B **75**, 224501 (2007).

13. G. Seyfarth, J. P. Brison, G. Knebel, D. Aoki, G. Lapertot, and J. Flouquet, Phys. Rev. Lett. 101, 046401 (2008).

14. S. -Z. Lin, J. Phys.: Condens. Matter **26**, 49 (2014).

15. A. Polyakov et al., Phys. Rev. B **85**, 245119 (2012).

16. P. Giannozzi et al., J. Phys. Condens. Matter **21**, 395502 (2009).

17. J. P. Perdew, J. A. Chevary, S. H. Vosko, K. A. Jackson, M. R. Pederson, D. J. Singh and C. Fiolhais Phys. Rev. B **46** 6671 (1992).

18. standard solid-state pseudopotentials library, https://www.materialscloud.org/discover/sssp/table/efficiency

19. P.M.C. Rourke and S.R. Julian, Comput. Phys. Commun. **183**, 324 (2012).

20. G, Seyfarth, J. P. Knebel, D. Aoki, G. Lapertot, and J. Flouquet, Phys. Rev. Lett. **101**, 046401 (2008).

21. M. A. Tanatar et al., Phys. Rev. Lett. **95**, 067002 (2005).

## Acknowledgements

We are grateful to Ilya Vekhter and Anton Vorontsov for numerous stimulating discussions. Theoretical work (SZL) was supported by DOE via LDRD program at LANL and in part, by the Center for Integrated Nanotechnologies, an Office of Science User Facility operated for the U.S. DOE Office of Science, under user proposals #2018BU0010 and #2018BU0083. Experimental work was conducted under the auspices of the U.S. Department of Energy, Office of Science, Basic Energy Sciences, Materials Sciences and Engineering Division.

## Figure captions

Fig 1. Schematic diagram of the thermal conductivity measurement in rotating magnetic field applied within the *a-b* plane of the tetragonal $CeCoIn_5$. The heat current was applied along the [100] axis. $\Theta$ is defined by the angle between the heat current and the direction of the magnetic field.

Fig 2. The magnetic field of 3 T was rotated by 180° from the crystallographic [0-10] axis to the [010] axis. The heat current was applied along the [100] axis. The dashed line is a fit to the sum of a 2-fold and a 4-fold terms. The dotted line is fit to a 2-fold term only.

Fig. 3. Thermal conductivity of CeCoIn$_5$ measured with rotating magnetic fields up to 7 T. The heat current was applied along the [100] axis. (a) The data for the low angle range, from -40° to 40°, highlighting the resonances at ±12°, with the sum of 2-, 4- and 8-fold fits to the background; (b) $\kappa/T$ vs. $T$ from (a) with the background fits subtracted; the resonances in field up to 5 T are clearly resolved; (c) $\kappa/T$ vs. $T$ in the regions of the high absolute value of angle, $\Theta < -50°$ and $\Theta > 50°$, with fits to the background obtained similarly to (a); (d) $\kappa/T$ vs. $T$ from (c) with the background fits subtracted, showing the persistent resonances features at $\Theta = \pm 78°$.

Fig. 4. Thermal conductivity of CeCoIn$_5$ measured with an in-plane rotating magnetic field up to 7 T, from Ref. [8]. The heat current was applied along the [110] axis. $\Theta$' is defined by the angle between heat current and magnetic field directions. (a) $\kappa/T$ vs. $T$ in the regions of the high absolute value of angle, -90 < $\Theta$' < 90°, highlights the resonances at ±33°, as marked by the vertical red arrows.

Fig 5. Three representative Fermi surfaces of CeCoIn$_5$ and the calculated density of state as a function of the direction of the magnetic field. The field is rotated from [100] ($\phi = 0°$) to [010] ($\phi = 0°$) within the $a$-$b$ plane.

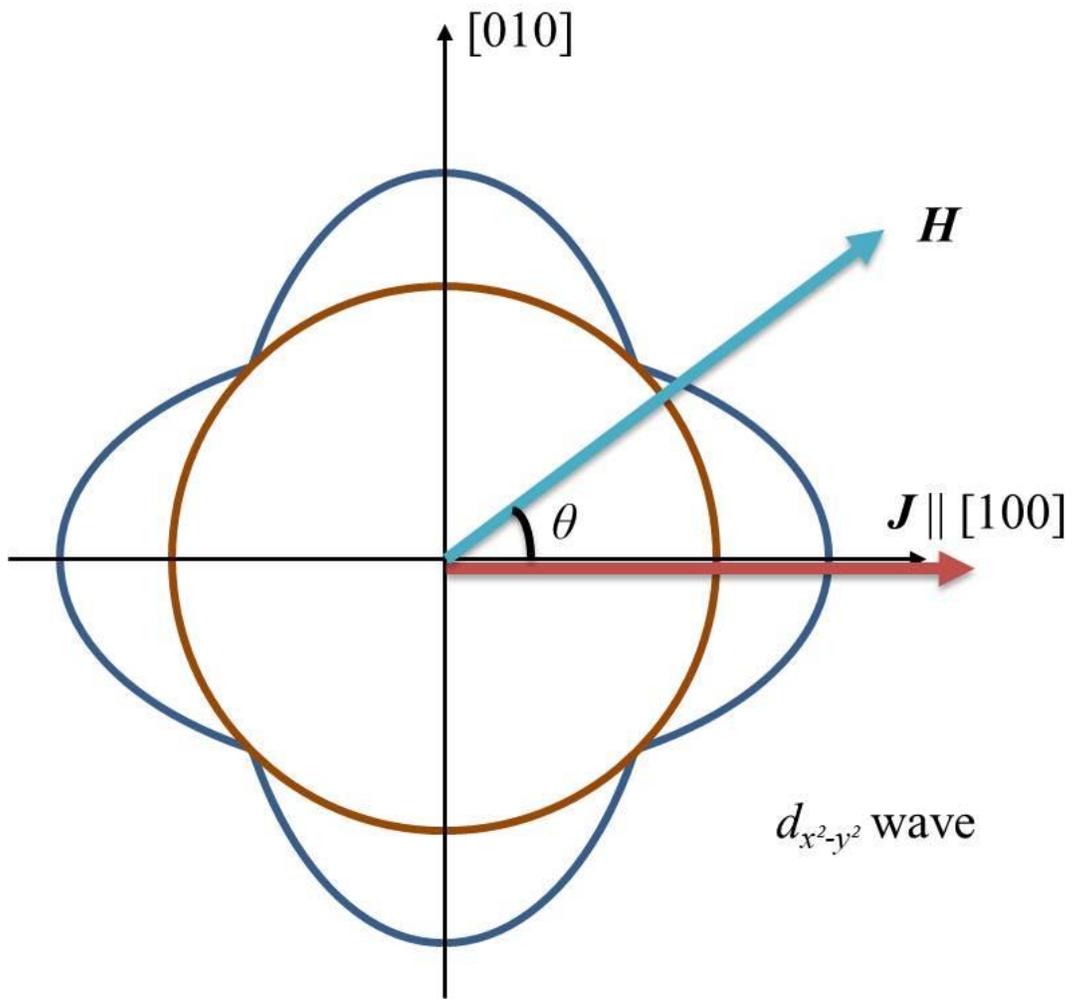

FIG.1

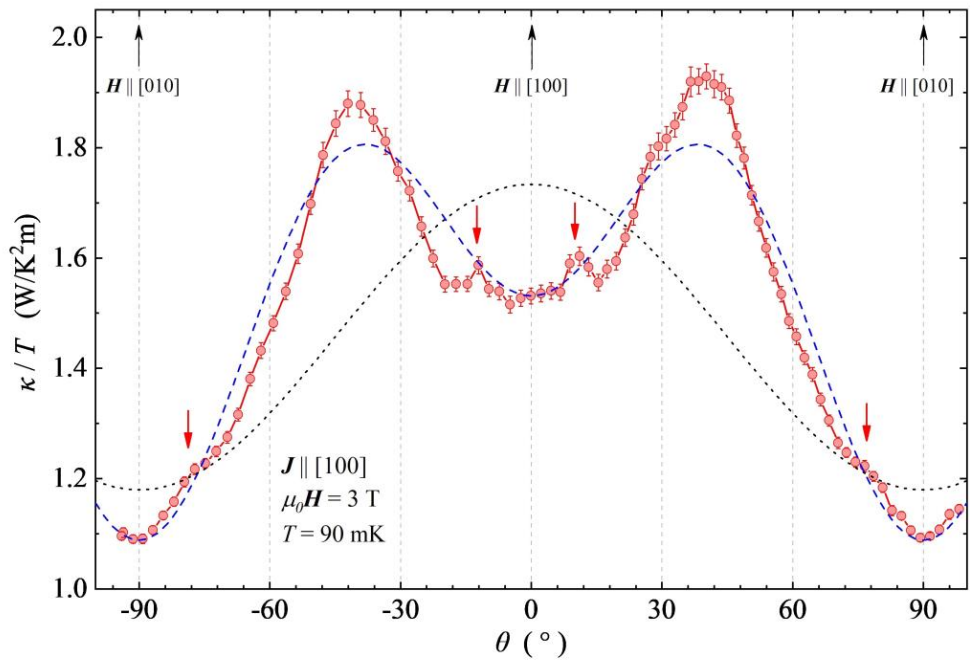

FIG.2

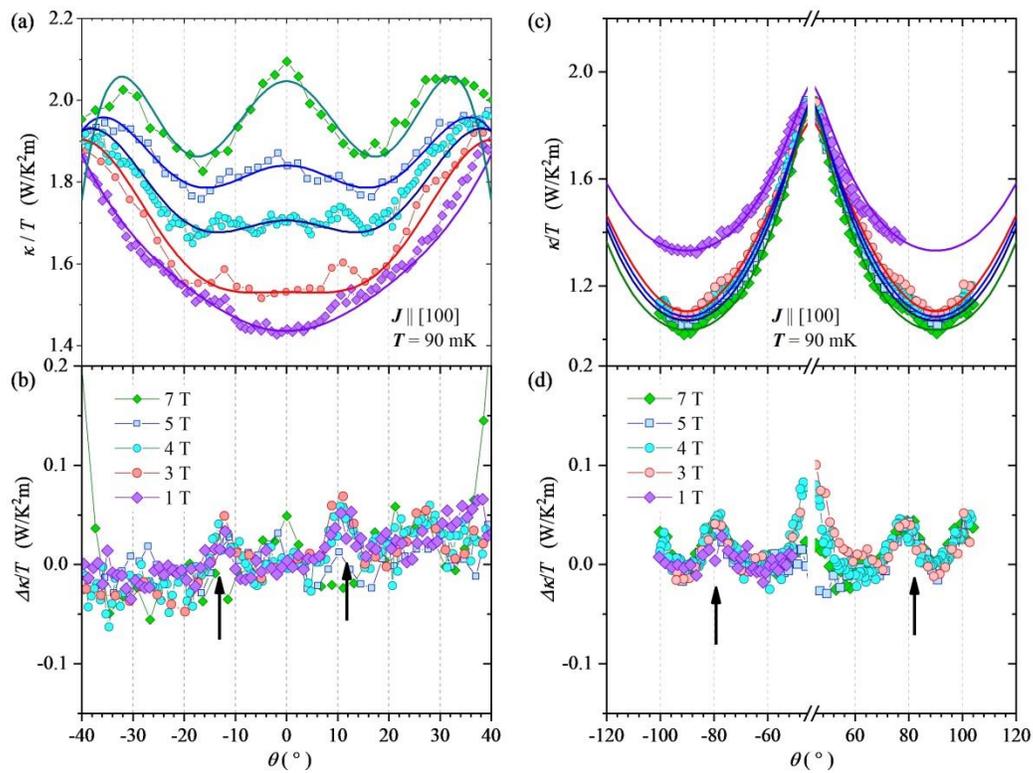

FIG.3

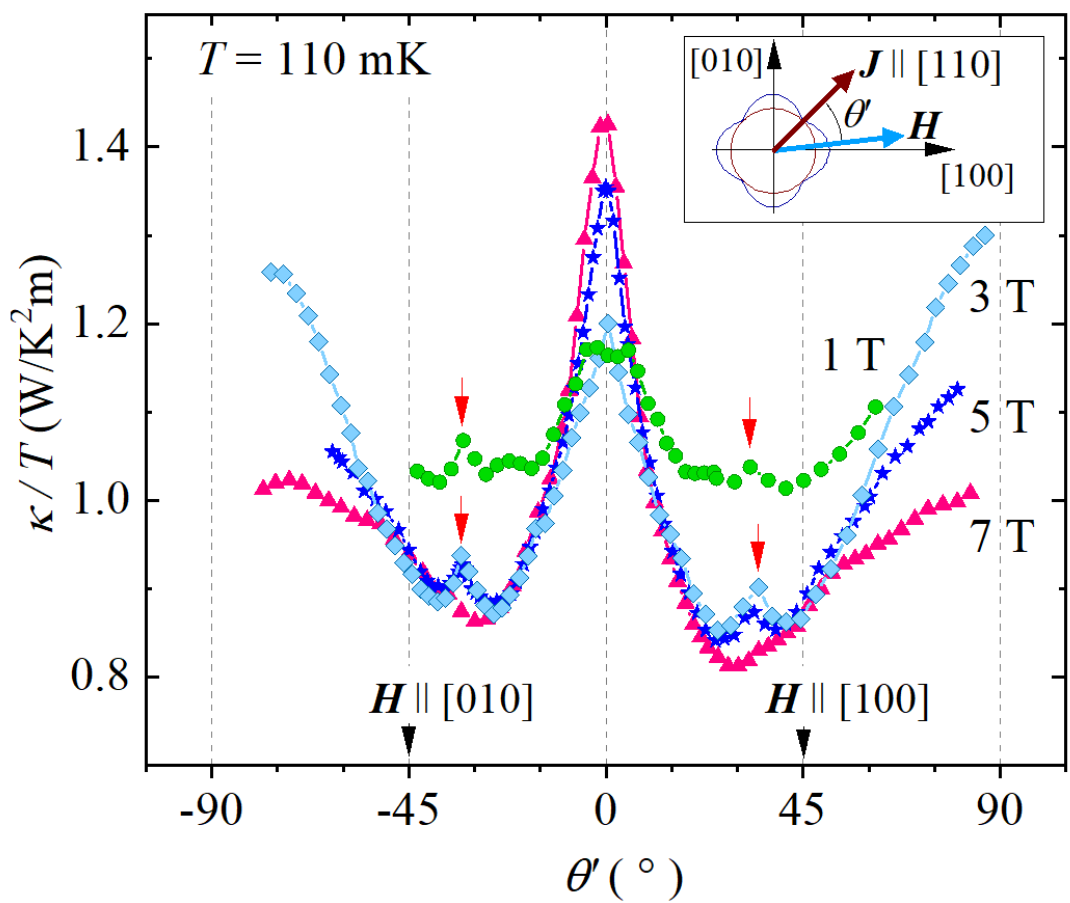

FIG.4

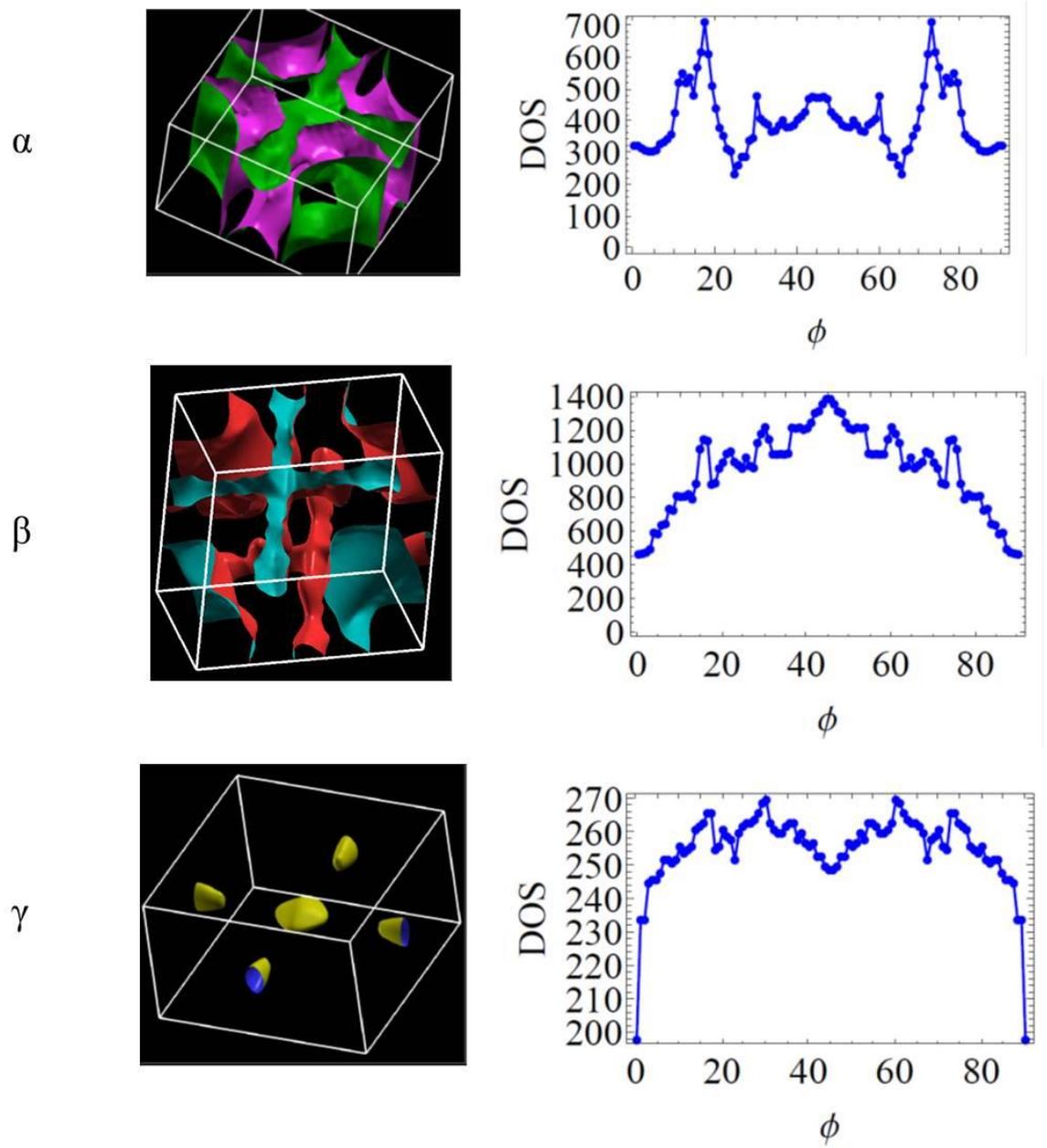

α

β

γ

FIG. 5

# Supplemental Material: Normal Fermi Surface in the Nodal Superconductor CeCoIn$_5$ Revealed via Thermal Conductivity

Sangyun Lee, Duk Y. Kim[*], Priscila F. S. Rosa, Eric D. Bauer,

Filip Ronning, J. D. Thompson, Shi-Zeng Lin[†], and Roman Movshovich[†]

Los Alamos National Laboratory, Los Alamos, New Mexico 87545, USA

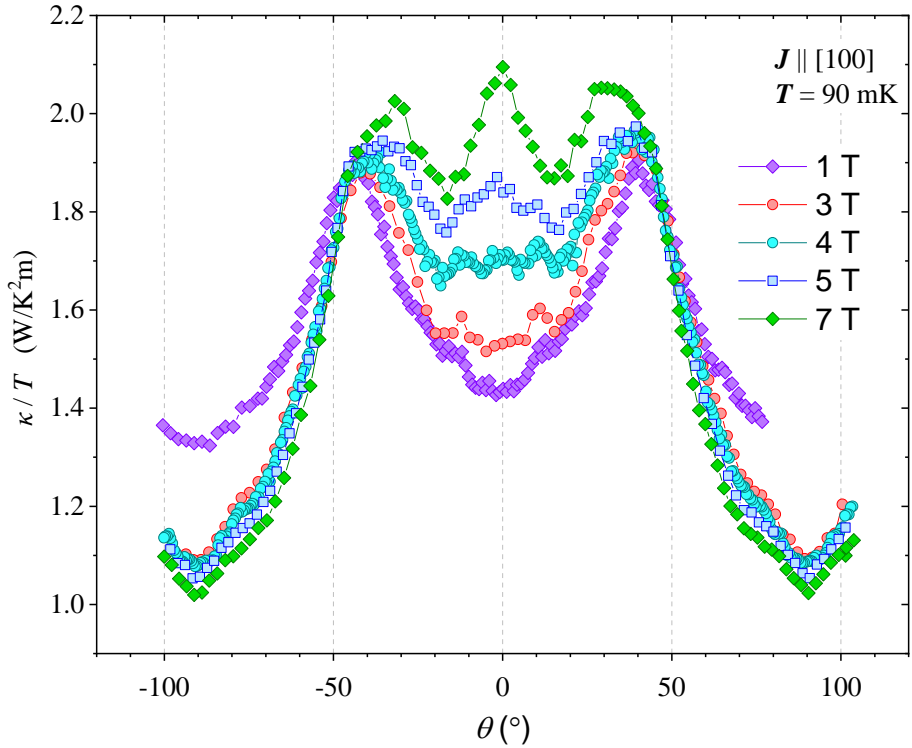

FIG. S1. The thermal conductivity of CeCoIn$_5$ divided by temperature $\kappa/T$ as a function of the angle between the rotating magnetic and the heat current J∥[100], for field between1 and 7 T. Small peaks are observed at ±12° and ±78°. The peak at 0º is due to the resonant vortex core contribution and rises with increasing magnetic field.